\documentstyle[aps,preprint]{revtex}

\tightenlines

\begin{document}

\title{Microwave detected, microwave-optical double
resonance of NH$_{3}$,
NH$_{2}$D, NHD$_{2}$, and ND$_{3}$:  \\I. Structure and force
field of the \~A
state.\thanks{This work is taken in part from the
Ph. D. thesis of Steven A. Henck.\protect\cite{Henck:1990}}}

\author{Steven A. Henck,\thanks{Present address: Texas Instruments, 13536 N.
Central Expy.%
, MS 992, Dallas, TX 75243}\   Martin A. Mason, Wen - Bin Yan,\thanks{Present
address: Energia, Inc., P. O. Box 1468, Princeton, NJ 08542}
and Kevin K. Lehmann}
\address{Department of Chemistry, Princeton University,
Princeton, NJ 08544}

\author{and Stephen L.\ Coy}
\address{Harrison Regional Spectroscopy Laboratory,
Massachusetts Institute of Technology, Cambridge, MA 02139}

\maketitle

\draft

\begin{abstract}

Microwave detected, microwave-optical double resonance was used to record the
\~A state electronic spectrum of NH$_{3}$, NH$_{2}$D, and NHD$_{2}$ with
both vibrational and rotational resolution.  To investigate ND$_{3}$ with the
same resolution as we had with our hydrogen containing isotopomers, a
strip-line
cell was constructed allowing the simultaneous passage of radio-frequency
and ultraviolet radiation.  Rotational constants were obtained as a function of
$\nu_{2}$ excitation and an \~A state equilibrium bond length was estimated at
1.055(8)~\AA. In addition, the harmonic force field for the
\~A state has been experimentally determined.  $f_{hh}$, $f_{\alpha \alpha
}-f_{\alpha \alpha '}$ ,
and $f_{rr}$ were found to be 1.06(4) ~aJ/\AA$^{2}$, 0.25(2) ~aJ, and 4.9
{}~aJ/\AA$^{2}$ respectively.
This calculated harmonic force field predicts that the asymmetry observed in
the NH$_{3}$ 2$^{4}$ band is due to a strong anharmonic interaction with the
4$^{3}$
level and the broad feature observed in the dispersed fluorescence spectrum
previously assigned to the 1$^{1}$ band is more likely attributable to the
4$^{2}$ level.

\end{abstract}


\clearpage

\section{Introduction}

The lowest excited singlet state of ammonia, the \~A state, has been subject to
extensive spectroscopic study by a range of methods.  However, since the first
report by Leifson\protect\cite{Liefson:1933}, it has been known that
conventional techniques
provide a limited amount of information.  Leifson reported
that the \~A state absorption begins at about 220 nm, principally consists of
a single long progression, and is diffuse.  He interpreted this diffuseness
as due to a rapid predissociation of the excited state such that the one photon
spectrum shows resolved vibrational but not rotational structure.  Walsh and
Warshop
\protect\cite{Walsh:1961} were able to show that the excited state is planar;
hence,
the long progression observed by
Leifson\cite{Liefson:1933}  resulted from a large displacement of $\nu_{2}$,
the out-of-plane bending mode.  They also showed that the \~A state has
A$_{2}^{''}$
electronic symmetry and an N--H bond length greater than that of the ground
electronic state.

With higher resolution than previous investigations,
Douglas\protect\cite{Douglas:1963}
 partially resolved the rotational structure in the two lowest vibronic bands
of
ND$_{3}$ and obtained rotational constants for these levels.
Analysis of the rotational constants neglecting vibration-rotation interactions
confirmed the prediction of Walsh and Warshop\protect\cite{Walsh:1961}
suggesting that
in the excited state, the \mbox{N--D} bond length was significantly displaced.
Such a large displacement in this coordinate also implied that in the
absorption spectrum,
a second progression in the symmetric stretching mode should also have been
observed.
As expected from above, a long progression in $\nu_{1}$ is observed in the \~A
state
resonance Raman spectrum\protect\cite{Ziegler:1984} and dispersed emission from
the
ND$_{3}$ 2$^{1}$ level\protect\cite{Gregory:1976}.

Several hypotheses have been given to explain how the $\nu_{1}$ progression
is missing in absorption but seen in emission.  As initially suggested by
Harshbarger\protect\cite{Harshbarger:1970}, one hypothesis proposes that the
separate
$\nu_{1}$ progression was not observed in absorption because the $\nu_{2}$ and
$\nu_{1}$ modes were in a near 3:1 degeneracy, the separate $\nu_{1}$
progression being buried underneath the more intense $\nu_{2}$ progression.
However,
even under the cold conditions of a jet expansion, no separate progression
was revealed for NH$_{3}$ or ND$_{3}$\protect\cite{Vaida:1987}.   Another one
proposes that the
$\nu_{1}$ mode predissociates so rapidly that its broad transition width
precludes its observation\protect\cite{Tang:1988}. Ashfold {\it et
al.}\protect\cite{Ashfold:1986a} have observed a
feature that they assigned to the $\nu_{1}$ fundamental in
dispersed fluorescence from the higher lying \~C$^{'}$ state.
They reported transition widths of ~500~cm$^{-1}$ for these bands.  In the
following paper\protect\cite{Henck:1994b},
 we will examine the predissociation of the \~A state in detail.  On
the basis of that work, we assign the barrier to N-H
dissociation as 2075~cm$^{-1}$, implying essentially direct
dissociation from levels with the N--H stretching mode
excited.  As a result, we believe that any $\nu_{1}$
progression would produce an essentially continuous
absorption.  Tang {\it et al.}\protect\cite{Tang:1990}  provided a third
explanation for
the lack of a $\nu_{1}$ progression in the absorption
spectrum.  Using a model surface, with a small change in
bond length at the ground state minimum, but a large change
near the excited state planar geometry, they calculate that
there should be $\nu_{1}$ activity in emission but not
absorption from the ground state.  Thus, the two need not
have the `symmetry' that is predicted by normal mode theory.

The \~A $\leftarrow$ \~X  transition has also been
thoroughly investigated by a variety of theoretical
techniques.  From the CASSCF and CEPA wavefunctions for the
symmetric modes, Rosmus {\it et al.}\protect\cite{Rosmus:1987} calculated the
potential
energy surfaces (PES) for both the \~X and \~A
states.  These calculations confirmed that the
\~A $\leftarrow$ \~X transition was accompanied by a
\hbox{planar $\leftarrow$ pyramidal} geometry change and that the N-H(D)
equilibrium
bond length is longer in the excited state; yet, the calculated \~A state N--
H(D)bond length was still much less than the experimental
estimate\protect\cite{Walsh:1961,Douglas:1963,Harshbarger:1970}.  They
proposed that this discrepancy was due to errors caused by
neglect of vibration-rotation interactions in reducing the observed
rotational constants to an equilibrium geometry.  The smaller displacement in
this coordinate reduces the expected intensity of the missing $\nu_{1}$
progression compared to the previous estimates\protect\cite{Harshbarger:1970}.

Imre and co-workers\protect\cite{Tang:1988,Tang:1990,Tang:1991}  have modeled
the \~A
$\leftarrow$ \~X absorption spectrum using a time-dependent formalism.
Using the unmodified potential of Rosmus {\it et al.}\protect\cite{Rosmus:1987}
 with
only the out-of-plane bend and symmetric stretching coordinates modeled, their
calculated spectrum peaks to the red of the observed one.  Agreement
between experimental and calculated spectra was only achieved by
modifying the potential of Rosmus {\it et al.}\protect\cite{Rosmus:1987}  to
include a longer
N-H bond length. However, increasing
the N-H bond length also increases the $\nu_{1}$ Franck-Condon
 factors.  Hence, a progression in $\nu_{1}$ is again predicted.  In their
work\protect\cite{Tang:1990},
 however, they found that $\nu_{1}$ and 3$\nu_{2}$ progressions moved in a
concerted
fashion to yield in absorption only a single progression.   Because of their
neglect of
the asymmetric stretching modes, their calculation did not include the
effects of predissociation to NH$_{2}$ and H, and therefore, the lifetime
broadening of these levels.

Rosmus {\it et al.}\protect\cite{Rosmus:1987}  calculated the $\nu_{1}$
frequency at
2835~cm$^{-1}$.  However, through dispersed fluorescence from the longer lived
\~C$^{'}$ state\protect\cite{Ashfold:1986b},  the frequency of this mode
was measured at 2300 $\pm$ 500~cm$^{-1}$\protect\cite{Ashfold:1986a}.
Dixon\protect\cite{Dixon:1988} resolved this discrepancy by including in
the theoretical determination interactions
between $\nu_{1}$ and the dissociative $\nu_3$ coordinate.
Inclusion of these interactions also enabled Dixon
to accurately model the observed transition widths\protect\cite{Dixon:1988}.
However, unlike the previous reports\protect\cite{Harshbarger:1970,Tang:1988},
a 1:3 degeneracy between $\nu_{1}$ and $\nu_{2}$ was not supported.

Utilizing the longer lifetime of the \~C$^{'}$
state\protect\cite{Ashfold:1986b},
  the \~A state has been studied by optical-optical
double resonance techniques (OODR)\protect\cite{Ashfold:1986a,Xie:1986}.
   These techniques simplify the spectrum by
being sensitive to changes in only a single rovibrational
level at a time.  In the OODR method, rovibrational selectivity is accomplished
by laser multiphoton excitation of a specific J$_{K}$ line in the
\~C$^{'}$ state.  Xie
and coworkers\protect\cite{Xie:1986},   who were the first to apply this method
to \~A state ammonia, used a second tunable laser to ionize
ammonia.  When this second laser was in resonance with a downward
\~C$^{'}$ $\rightarrow$ \~A transition from the initially populated level, the
ionization signal was depleted.  They reported the NH$_{3}$ \~A state
2$^{0}$, 2$^{1}$, and 2$^{2}$ bands with both J and K resolution.
 They determined the B and C rotational constants for these levels to 0.3-0.03%
{}~cm$^{-1}$ accuracy.  Interestingly,
these constants agree within experimental error with
previous estimates from line shape analysis\protect\cite{Ziegler:1985}.
 Later, Ashfold {\it et al.}\protect\cite{Ashfold:1986a} performed a similar
experiment except with fluorescence-dip detection.  They used much lower
dump laser power and observed a significant decrease in
transition widths demonstrating that the results of Xie {\it et
al.}\protect\cite{Xie:1986}  suffered
from significant power broadening.

Microwave detected, microwave-optical double resonance (MODR) has
been used to investigate the vibrational overtones of
ammonia in the visible region\protect\cite{Coy:1986}.   As the OODR techniques
previously described, this technique allows one to measure the absorption
spectrum of a single rovibrational level
at a time.  The principal advantage of this method is that
selectivity is based upon the well-known ground state microwave
spectrum\protect\cite{Sharbaugh:1949,Good:1947,Townes:1955}.
In contrast, the OODR technique uses the less characterized
\~C$^{'}$ state transitions\protect\cite{Ashfold:1986b}.   Furthermore,
predissociation in the
\~C$^{'}$ state limits the range of \~A state rovibronic levels that can be
reached.  Since the ground state microwave transitions are known for all
ammonia isotopomers\protect\cite{Hermann:1958,DeLucia:1975,Fusina:1985},
  the MODR method can be
relatively easily extended to the investigation of the entire isotopomeric
series.  Following the completion of this work, Endo and
coworkers\protect\cite{Endo:1990}
  reported the use of this technique to the study of the 2$^{1}$ level of the
NH$_{3}$ \~A state.  They reported the homogeneous linewidths and
positions of nearly fifty transitions to this band.  Our work, even though not
as extensive as theirs for this level, confirms their findings.  We have
extended this technique to the study of the higher NH$_{3}$ \~A state
vibronic levels.  In addition, we have applied this technique to the
study of the isotopomers NH$_{2}$D, NHD$_{2}$, and ND$_{3}$.

In ND$_{3}$, the inversion
frequency drops to less than 2 GHz\protect\cite{Hermann:1958,Fusina:1985} so
that
for the investigation of this isotopomer, the identical
spectrometer could not
be used.  In order to investigate the double resonance spectrum of
this isotopomer, we constructed a radio-frequency detected, radio-frequency
optical double resonance spectrometer (RFODR) similar to the MODR
spectrometer.  The principal difference between the two instruments was that in
the RFODR spectrometer, the waveguide cell was replaced by a strip-line
 cell that simultaneously allowed the passage of radio-frequency and
ultraviolet radiation.  Using this cell, we were able to record the
rovibrational contours of the 2$^{0}$, 2$^{1}$, and 2$^{2}$ bands of the
ND$_{3}$
\~A state with the same sensitivity that we had with the other isotopomers.

In this paper, we report on the \~A transition frequencies we have observed
by MODR and RFODR for the entire isotopomeric series.
This experiment represents the most extensive
spectroscopic investigation of the \~A state of ammonia conducted to date.
The goals of
this study were the following.  First, using the RFODR
spectrometer, the resolution in the ND$_{3}$ 2$^{1}$ spectrum was increased
such
that the centrifugal distortion coefficients for this level were derived.
Second, for all isotopomers, precise rotational constants and vibrational
term values were derived as a function of $\nu_{2}$.  From this data set, the
harmonic force field of the \~A state was determined.  This newly derived force
field
leads us to reassign the broad features observed by Ashfold {\it et al.}%
\protect\cite{Ashfold:1986b} in dispersed fluorescence from the \~C$^{'}$ state
as due to $\nu_4$ instead of $\nu_{1}$.
Third, with the more extensive rotational data, the
vibration-rotation interaction term values were estimated allowing corrections
to be obtained for the experimentally determined N-H(D) equilibrium bond
length.  This newly derived experimental estimate for $r_{e}$ is in excellent
agreement with the {\it ab initio} prediction of Rosmus {\it et
al.}\protect\cite{Rosmus:1987}.
And lastly, these MODR experiments have allowed us to more accurately model
the rotational and vibrational
dependence of the transition widths.  This topic will be
addressed in the following paper\protect\cite{Henck:1994b}.

\section{Experimental}

A schematic of the MODR spectrometer is shown in
figure~\protect\ref{Schematic}.
This apparatus is similar to that used by Coy and
Lehmann\protect\cite{Coy:1986}
except that ours does not include a microwave bridge to null the detector.  The
bridge increased signal to noise in the earlier work since it allowed for the
use of greater ammonia pressure.  In the current experiment, ammonia
pressure was limited by the strong absorption cross section of the \~A
$\leftarrow$ \~X transition and the need to keep the sample
optically thin so as not to distort the observed lineshapes.  The pulsed
optical radiation was provided by a XeCl excimer pumped dye laser (Questek
2420/
Lambda Physik 3002E, respectively) using one of the following two schemes.
To generate radiation from 207 nm to 220 nm, output from the dye laser
using stilbene 3 (Exciton) as a gain medium was frequency-doubled in a
$\beta$-BBO
(Skytek) cut at $80^{\circ}$; or, to generate radiation from 197 to 210 nm, the
output of rhodamine B (Lambda Physik) was tripled.  Tripling was accomplished
by
first doubling in a KDP crystal using an Inrad Autotracker II.  The remaining
fundamental and second harmonic (~300 nm) output of the KDP
crystal are orthogonally polarized.  To rotate the two beams such that they
are parallel polarized, they are both passed through a 600
nm 0-order half-wave plate that acts as  an approximate full waveplate for
the second harmonic.  The resulting nearly parallel polarized beams are
then sum frequency mixed in the type 1 $\beta$-BBO crystal.  Typical
ultraviolet
power was 50-100 $\mu$J per pulse.

In the MODR experiments, continuous-wave microwave radiation
was provided by a series of Oki klystrons phaselocked at
selected ground state inversion frequencies.  A two meter long WR-42 (``P''
band)
waveguide cell was used.  At each end of the cell, a 0.020$^{''}$ (0.51 mm)
thick
sapphire window and a specially machined E-plane bend were mounted.  These
bends
had a 1/8$^{''}$ hole drilled at the zero current line to allow ultraviolet
radiation to enter and leave the cell, while the microwave radiation turned the
corner.  The cell was internally gold-plated to significantly reduce ammonia
adsorption.  The microwave power transmitted through the cell was detected by
a Schottky-barrier detector diode and recorded by gated boxcar
integrator triggered by the pulsed laser.  The ultraviolet and microwave
radiation had
parallel polarizations.  Since the microwave probes were Q branch
rotation-inversion transitions, polarization effects enhance Q branch
transitions, and weaken P and R branch transitions compared to the
H\"onl-London line
strength factors applicable to absorption from an isotropic sample.

NH$_{3}$(Matheson, 99.995\%) and ND$_{3}$ (Cambridge
Isotopes, 99.5\%) were used after purification through several freeze pump thaw
cycles.  The gas samples were
flowed continuously through the cell at a set pressure to avoid
depletion of the observed signals caused by photodecomposition.  The mixed
isotopes were prepared by combining NH$_{3}$ and ND$_{3}$ in the
ratios 1:2 and 2:1 to maximize the ratio of NHD$_{2}$ and NH$_{2}$D,
respectively. That the gas samples consisted of a mixture of isotopomers
was not a problem since a transition between a single ground state inversion
level of a particular isotopomer was selected and monitored by the microwave
radiation.  In all experiments, optimal signals were obtained at pressures
between 10 and 70 mtorr.

Scanning and calibration of the laser were achieved by
conventional methods\protect\cite{Mason:1993}.    Briefly, the transmission of
an etalon and the
optogalvanic signal from a Fe/Ne\protect\cite{Phelps:1982} hollow cathode lamp
or
the absorption of a $^{131}$Te$_{2}$ cell\protect\cite{Cariou:1980}  were
recorded for
relative and absolute frequency calibration.
The ultraviolet power was monitored for both signal normalization, and for
computer control of the $\beta$-BBO crystal phase matching angle.  At the
same time, the detected microwave signal was amplified and sent to a gated
boxcar integrator.  These four signals were recorded by an IBM AT that also
stepped the dye laser and maximized the ultraviolet conversion.

When the ultraviolet frequency coincided with a \~A $\leftarrow$
\~X transition, the optical pulse created a transient population
difference between ground state inversion levels.  The microwave signal is
absorptive
if the upper state is dpleted by the optical pulse, and emissive if
the lower state is depleted.  This is determined by the selectrion
rules for the \~A$\leftarrow$\~X transition.
  These rules are that symmetric inversion levels
have allowed transitions to even number of quanta in $\nu_{2}$ while
anti-symmetric
 inversion levels have allowed transitions to odd number of quanta in
$\nu_{2}$.  Hence, the detected microwave intensity initially increased or
decreased depending whether the final $\nu_{2}$ quantum number was even or odd,
respectively.  A typical double resonance signal is shown in
figure~\protect\ref{Nutation}.  Since
this signal was observed through the excitation to the NH$_{2}$D 2$^{0}$ level,
the microwave intensity initially increased.  Phase sensitivity was
achieved by setting a boxcar gate over the first hump of this Rabi nutation.

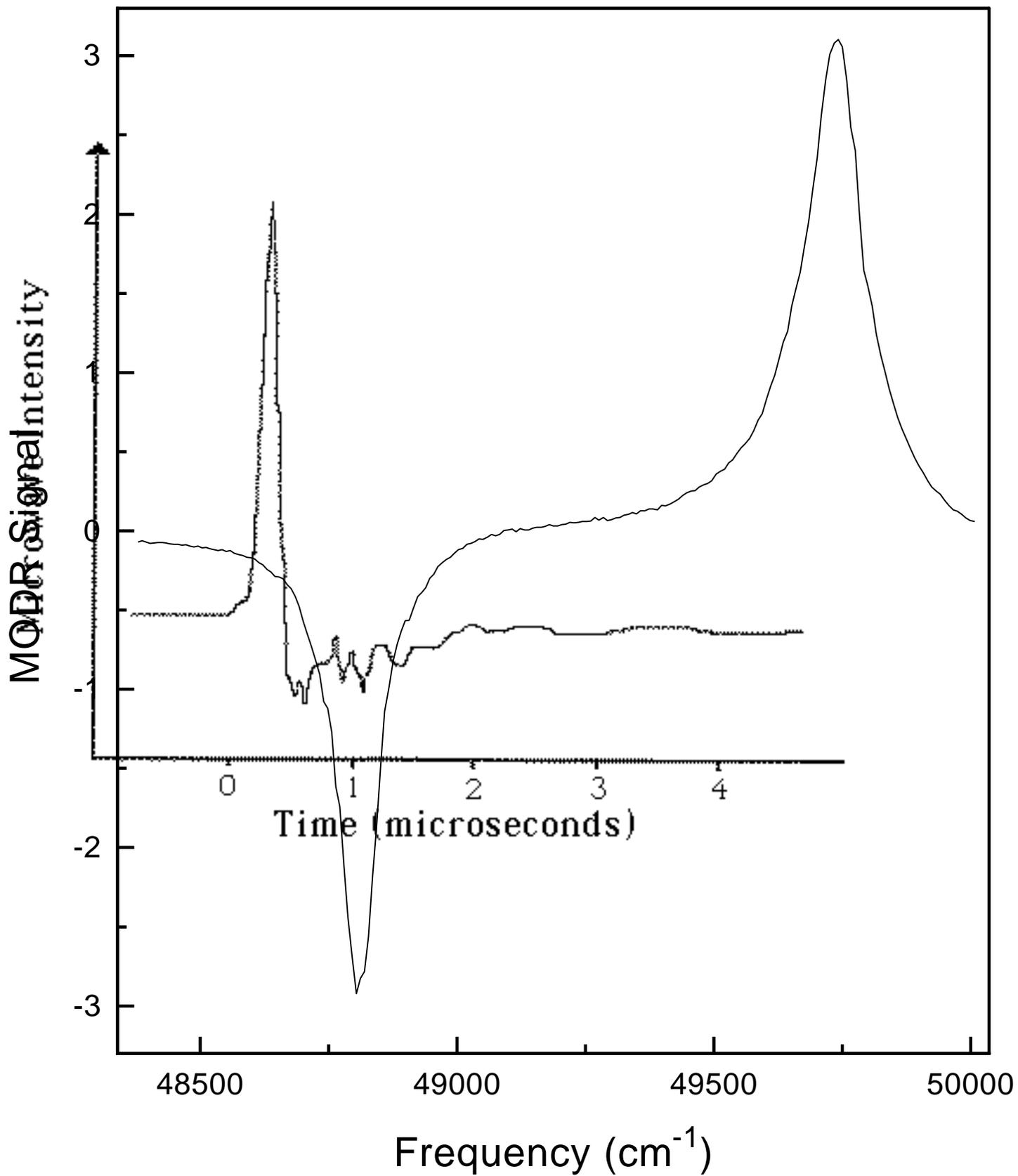
\begin{figure}
\caption{
Observed MODR signal to the 2$^{3}$ and 2$^{4}$ levels in the NH$_{3}$
\~A state with the microwave frequency locked on
the ground state 6$_{6}$ inversion doublet.  The parallel
polarization
of the two fields and Honl-London factors permit observation
of
only the Q branches.}
\label{Representation}
\end{figure}

A representative MODR scan is shown in figure~\protect\ref{Representation}.
The locked
microwave frequency was resonant with the NH$_{3}$ J$_{K}$ = 6$_{6}$
inversion doublet
as the optical radiation was scanned.  When the ultraviolet
frequency was in
resonance with the 6$_{6}$ line of the \~A 2$^{3}$ $\leftarrow$ \~X
2$^{0}$ band, the upper component of the inversion doublet was
depleted and a decrease in microwave intensity detected.  Further to the
blue when the ultraviolet frequency came into resonance with the 6$_{6}$ line
of the \~A 2$^{4}$ $\leftarrow$ \~X 2$^{0}$ band, the lower component of the
inversion doublet was depleted and an increase in microwave intensity was
detected.  For this particular scan, only the Q branches
were observed due to the parallel polarization of the two fields and the
H\"onl-
London factors, both of which enhance the Q over the R branch transition
in this case.

The inversion transitions in ND$_{3}$ are near 1.6
GHz\protect\cite{Hermann:1958,Fusina:1985},  well below the
cut off frequency of our MODR cell.  While in principle, one
could build a cell using much larger waveguide, this would necessarily lead to
a very poor overlap of the microwave radiation with the ultraviolet
pumped molecules.  Instead, we constructed a RFODR cell based upon a strip-line
cell.  A cross section of the strip-line cell is shown in
figure~\protect\ref{Strip_Line}.  The
main components of the cell are a 5 $\times$ 5 $\times$ 200 cm aluminum outer
body,
 a 0.8 $\times$ 0.8 $\times$ 200 cm aluminum center conductor,
an aluminum ground plane, and two S1-UV windows mounted on flanges at each end
of the
cell.  The center conductor was held at a fixed distance from the ground plane
by KEL-F blocks. Originally, the distance (~2 mm) was chosen to reduce
radio-frequency
reflections by impedance matching the cell to the source and
detector impedance.  However, with this spacing, beam divergence and
scattering of the ultraviolet light produced a large photoelectron signal
from the aluminum strip and ground plane that prevented detection of
any double resonance signal.  By placing rf hi-pass filters at each end of
the cell, adding approximately 100 mtorr SF$_{6}$ that acted as an electron
scavenger,
and doubling the spacing between the center conductor and the ground plane, the
anomalous photoelectron signal was reduced to a manageable level.  No
increase in noise level due to impedance mismatch was noticed.
Radiation for the RFODR spectrometer was produced by a
General Radio 1218-A unit oscillator. The rf power was measured by
a HP 8473B crystal detector while all other features of the spectrometer
are identical to the MODR setup previously described.

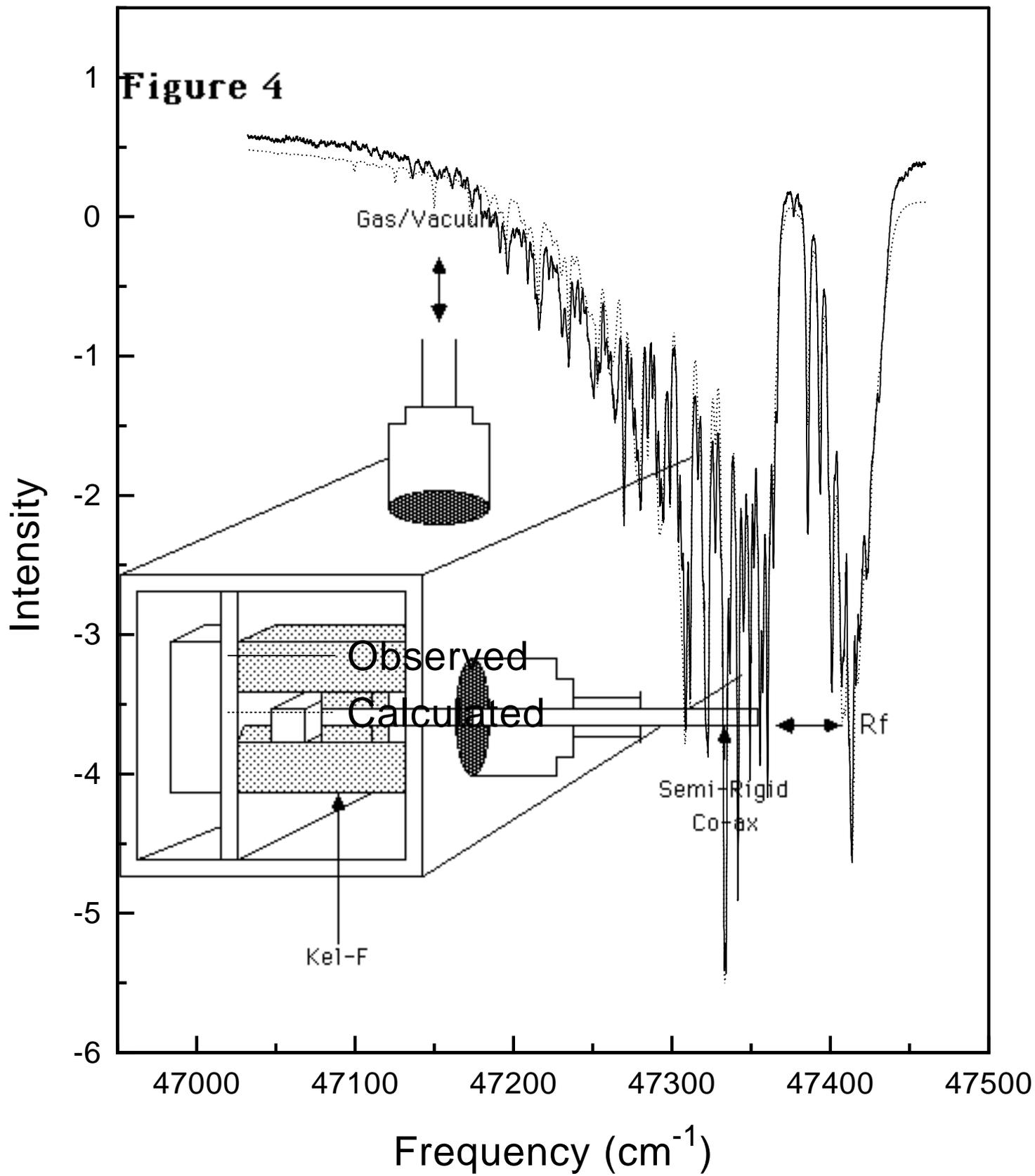
\begin{figure}

\caption{Calculated and observed fluorescence excitation spectrum
of the ND$_{3}$ 2$^{1}$ band.}
\label{ND3LIF}

\end{figure}

As will be discussed below, the lifetime for predissociation
from the lowest levels of ND$_{3}$ is much longer than those of the
other isotopomers.  As a result, as first shown by
Douglas\protect\cite{Douglas:1963},
  one observes partial rotational structure even in absorption.  The double
resonance
  data was dominated by high K transitions since the intensity of the
radio-frequency probe
transition scales as (K/J(J+1))$^{2}$.  A fluorescence excitation spectrum of
the
ND$_{3}$ \~A state 2$^{1}$ band was taken since the P and R branches have the
highest H\"onl-London factors for the low K levels.  The pulsed laser
system described in the double resonance experiments was used as the excitation
source.  Calibration was achieved in an identical fashion.  The output was
passed
through a 34 cm long, heavily baffled cell in which ND$_{3}$ was slowly
flowed to maintain a pressure of 0.5 torr.  Fluorescence was detected at right
angles to the laser propagation direction with an EMI 9635QB
photomultiplier and boxcar detected.  The resulting
spectrum is shown in figure~\protect\ref{ND3LIF}.

\section{Data Analysis}

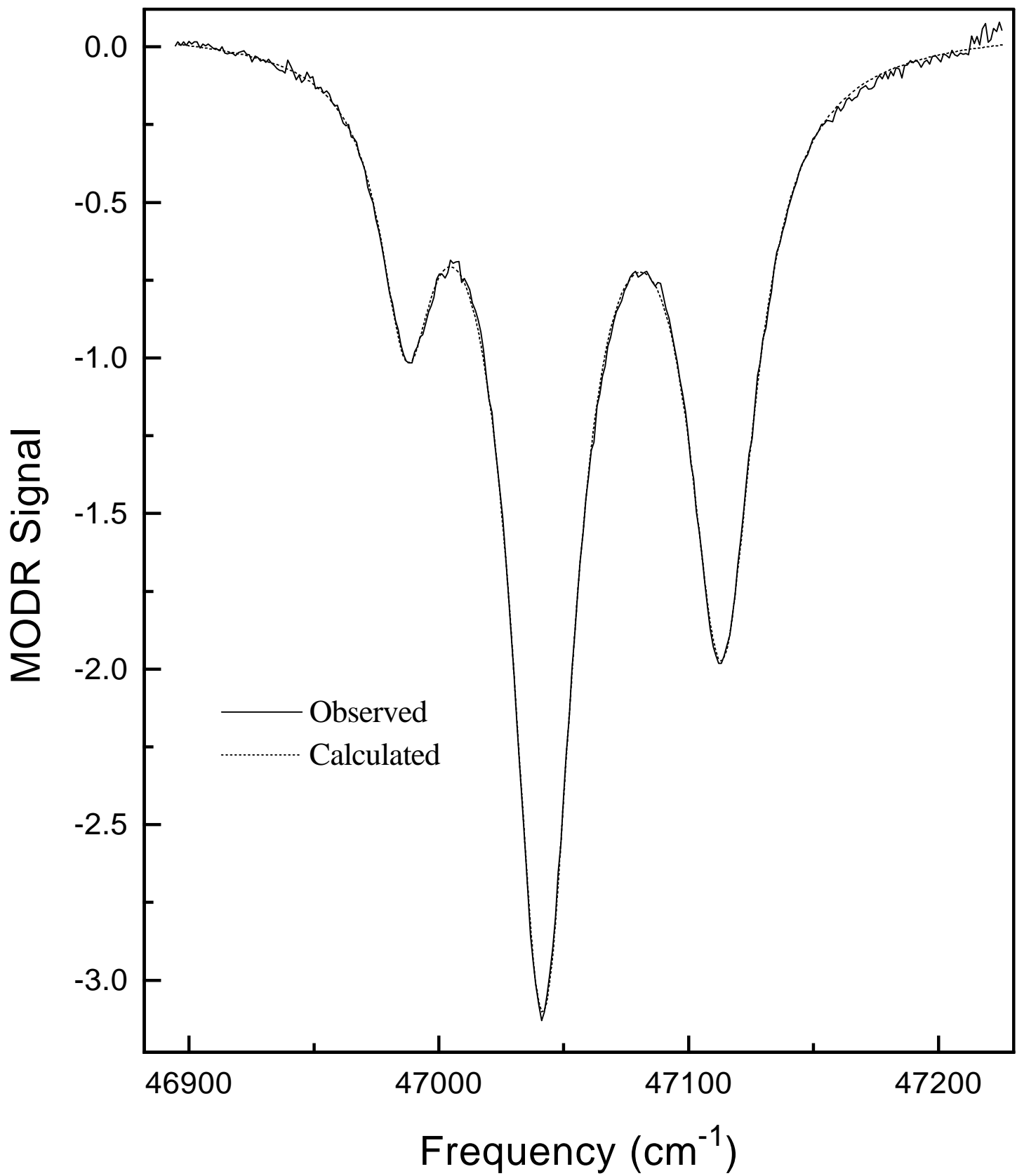
\begin{figure}

\caption{The observed and calculated transitions for the J$_{K}$ = 3$_{2}$
rotational level
in the NH$_{3}$ 2$^{1}$ band.}
\label{Fits}

\end{figure}

The observed peaks were fit by a sum of Lorentzians to
determine the height, width, and position of each spectral feature.  All
transitions were adequately described by a single Lorentzian accept those
observed in the NH$_{3}$ 2$^{4}$ band.  This system demonstrated a regular
asymmetry that prevented a quantitative treatment of these transitions.  The
transition
frequencies determined from the fits are given in
tables \protect\ref{NH3Freqs}, \protect\ref{NH2DFreqs},
\protect\ref{NHD2Freqs}, and \protect\ref{ND3Freqs}.  As an
example of the quality of the fits, in figure~\protect\ref{Fits} are presented
the calculated and
observed transitions for the J$_{K}$ = 3$_{2}$ rotational level in the
NH$_{3}$ 2$^{1}$ band. The observed widths are
much larger than the laser bandwidth (0.4~cm$^{-1}$) or the
Doppler width (0.1~cm$^{-1}$) and are therefore, a direct measure of the
predissociation
lifetime.  These widths as they relate to the predissociation rates will be
presented and discussed in the following paper\protect\cite{Henck:1994b}.

The transition frequencies for the pure isotopomers, NH$_{3}$ and ND$_{3}$,
were fit to a rigid symmetric top spectrum constraining the ground state
constants to the literature values.  In these fits, each observed transition
frequency was weighted by the squared uncertainty from the lineshape fits.
However, some of the lineshape fits yielded
unreasonably small error estimates causing these transitions to be
weighted too heavily.  To correct this problem, a constant term that reflected
systematic errors was added to the fit uncertainty for each transition
frequency
included.  This constant was chosen to equate the $\chi^{2}$ for the fit to
the number of degrees of freedom.  Fits were then redone with the new
weights, and the process iterated until convergence.  The rotational
constants determined by this procedure are presented in
table \protect\ref{HFits} and table \protect\ref{DFits} along
with previous determinations.  Where a comparison can be made, our results
agree favorably with other
reports\protect\cite{Douglas:1963,Ziegler:1984,Xie:1986}.
Additionally, two more points
should be noted.  First, we determine C better than the other rotational
constants due to the predominance of high K levels in our data set.  And
second,
the broad transition widths  of the hydrogen containing isotopomers
limited the accuracy of the experimental transition frequencies.
Therefore, in these cases, distortion constants were not included in these
fits.

One of the goals of the present work was the determination of the
centrifugal distortion constants for the lowest two ND$_{3}$ vibronic
bands.  At first, a six parameter fit to determine the three
distortion constants, D$_{J^{2}}$, D$_{JK}$, and D$_{K^{2}}$, was performed;
results from these fits are given
in column 1 of table~\ref{DFits}.  The parameters from the fits
yielded unphysical values and were moderately correlated.  Thus, utilizing
relationships among the distortion constants for a planar symmetric top, a
five parameter fit to $\tau_{xxxx}$ and $\tau_{zzzz}$ was also
performed and results from these fits are reproduced in column 2 of
table~\ref{DFits}.  A four parameter fit where
$\tau_{zzzz}$ was constrained to zero was also performed for the
2$^{1}$ level as shown in column 3 of table~\ref{DFits}.  Results from
these fits will be further elaborated in the discussion section.

The observed inertial defects for the NH$_{3}$ and ND$_{3}$ vibronic levels is
given in table~\protect\ref{Defects}.  These quantities were fit to
$\Delta_{obs}=\Delta_{o} +\Delta_{\nu_{2}}(\nu_{2}+1/2)$
to determine the dependence on $\nu_{2}$ excitation.  The determined values
for $\Delta_{o}$ and $\Delta_{\nu_{2}}$ are also presented in
table~\protect\ref{Defects}.  These values were used in the fits of the
mixed isotopomer data as discussed below.

The transition frequencies for the mixed isotopomers was fit to a rigid
asymmetric top spectrum.  Due to the limited number of observed transitions for
these two species, all three rotational constants could not be determined
independently.  To improve these fits, we constrain the calculated inertial
defect for the two mixed isotopomers to lie between the values observed for
the two pure isotopomers.  As an example for the NH$_{2}$ 2$^{1}$ level, we
constrain the inertial defect to the following,
\begin{equation}
\Delta_{mixed} = \frac{2\Delta(NH_{3},\nu_{2}=1)+\Delta(ND_{3},\nu_{2}=1)}{3}.
\end{equation}
B and C were independently varied while A was constrained to reproduce
this calculated inertial defect.  The determined values from
these fits are presented in table~\protect\ref{HFits}.

{}From the fit of the RFODR data on ND$_{3}$, we obtained preliminary
rotational constants for the 2$^{1}$ band.  Using these
constants, including estimates for the J, K dependence of the width to be
discussed in the following paper\protect\cite{Henck:1994b},
the fluorescence excitation spectrum of the
2$^{1}$ band was simulated.
This procedure allowed assignment of many of the strong features in
the excitation spectrum, particularly the P branch.  These assigned lines
were used in the final refinement of the rotational constants of the 2$^{1}$
band.
The simulation of the 2$^{1}$ band, using the final constants, is overlaid with
the
experimental spectrum in figure~\protect\ref{ND3LIF}.

\section{Discussion}

In the previous \~A state ammonia work, there have been
substantial differences in the
experimental\protect\cite{Walsh:1961,Douglas:1963} and
theoretical\protect\cite{Tang:1988,Rosmus:1987}
determinations of the \~A state equilibrium bond length.
The resolution of this controversy was one of the primary goals
of the present work.  Using the MODR technique, rotational constants were
obtained for all the vibronic bands studied.  Given the $D_{3h}$ planar
equilibrium structure of the \~A state\protect\cite{Walsh:1961},  there is only
one
structural parameter, the N--H equilibrium bond length, which in principle
could be
determined from the B or C rotational constants.  The effective bond lengths
derived
through the B rotational constant data varied considerably, likely due to
significant vibration - rotation interactions induced by b axis
rotation.  The effective bond
lengths obtained from the C rotational constants showed less
variation.  Hence, we have estimated the equilibrium bond lengths by
attempting to correct for the smaller vibration-rotation interactions in
the observed C rotational constants.

For the two pure isotopomers, we used the fitted
rotational constants in the $\nu_{2}$=0 and $\nu_{2}$=1 vibronic levels to
calculate $\alpha_{2}^{C}$.  For NH$_{3}$, $\alpha_{2}^{C}$ was determined to
be
-0.07(5)~cm$^{-1}$ while
for ND$_{3}$ it was -0.036(8)~cm$^{-1}$.  These values were used
to correct for vibration-rotation effects caused by $\nu_{2}$ excitation.
After
correction for this motion, an averaged bond length, $r_{o}$, was calculated
for both species.  For NH$_{3}$, $r_{o}$ was
determined to be 1.082(3) \AA\ while for ND$_{3}$ it was 1.074(1) \AA.  For the
remaining modes, $r_{o} - r_{e}$ should essentially scale as the
inverse square root of the reduced mass\protect\cite{Laurie:1958}. For these
two bands,
the effects of the bend-stretch Fermi resonance (which is responsible for
the rapid rise in the dissociation rate for
levels with $\nu_{2} > 1$) does not play a significant role.  Thus, assuming
the
reduced mass of the remaining modes is dominated by H(D), we can
extrapolate the above $r_{o}$ values for NH$_{3}$ and ND$_{3}$ to yield a value
for
$r_{e}$ of 1.055(8)~\AA.  This result is in quantitative agreement with the
{\it ab initio}
 value determined by Rosmus and co-workers\protect\cite{Rosmus:1987},  and
confirms their
suggestion that the apparent disagreement
between theory and experiment was due to the neglect of vibration-rotation
interactions in deducing the bond length from the accurately determined
rotational constants of Douglas\protect\cite{Douglas:1963}.    It is
interesting to
note that Douglas pointed out that his estimated bond length had an unknown
uncertainty, but this caveat was dropped in almost all subsequent references to
this `experimental' bond length.

Another goal of the present work was the experimental
determination of the \~A state harmonic force constants.  There exist
{\it ab initio} estimates of a few of these
constants\protect\cite{Rosmus:1987,McCarty:1987},
  but none have been experimentally determined.  Due to the redundancy
condition on
  the in-plane angles, the in-plane force field is described by only four
constants,
$f_{rr}$, $f_{rr'}$, $f_{\alpha\alpha }-f_{\alpha \alpha '}$, and $f_{\alpha
r}-f_{\alpha r'}$.  The out-of-plane force field is described by a single
force constant, $f_{hh}$.  To
first order, two of these force constants, $f_{rr'}$ and $f_{\alpha
r}-f_{\alpha r'}$,
would only split the mode frequencies.  Furthermore, assuming these constants
to
be reasonably close to the values of the electronic ground
state\protect\cite{Hargiss:1988},
we anticipate these to be small and have little effect on the observed
frequencies.
In fact, comparison of the Q$_{1}$ and r(NH) {\it ab initio\/} potentials of
McCarthy
{\it et al.}\protect\cite{McCarty:1987} predict that $f_{rr'}$ is
less than 0.1 ~aJ/\AA$^{2}$.  Initially, we take these force constants to be
zero leaving
only three force constants to be determined.

We had hoped that we could use the centrifugal distortion constants to
help determine the harmonic force field.  For both the ND$_{3}$ 2$^{0}$ and
2$^{1}$ bands,
we have accurately determined the rotational constants as well as the
centrifugal distortion constants, $D_{J}$, $D_{JK}$, and $D_{K}$.
Unfortunately, we were unable to use the same procedure to determine the
centrifugal distortion constants for the 2$^{2}$ band due to the larger
observed
transition widths.  We note that for a planar symmetric top, the centrifugal
distortion constants are related by\protect\cite{Gordy:1984}
\begin{equation}
D_{JK}=-\frac{2}{3}(D_{J}+2D_{K}).
\end{equation}
The fits were constrained to obey this relationship.  The centrifugal
distortion
constants are related to the force constants by\protect\cite{Gordy:1984}:
\begin{equation}
\tau_{xxxx}=-4D_{J}=-\frac{1}{2}(4 \pi B)^{4}\sum_{i,j} [J_{xx}^{(i)}]_{e}
f_{ij}^{-1}[J_{xx}^{(j)}]_{e}
\label{taux}
\end{equation}
\begin{equation}
\tau_{zzzz}=\frac{4}{3}(D_{K}-D_{J})=\frac{1}{2}(4\pi C)^{4}
(\frac{2}{3}mr^{2})^{2}f_{rr}^{-1}
\label{tauz}
\end{equation}
where $[J_{xx}^{(i)}]_{e}=\delta I_{xx}/\delta Q_{i}$  at the
equilibrium geometry. Our observed transition frequencies were then used to
determine $\tau_{xxxx}$  and $\tau_{zzzz}$ for all three bands.  These values
are also
included in table~\protect\ref{DFits}.

The $\tau_{xxxx}$ values listed in table~\protect\ref{DFits} for the 2$^{1}$
and 2$^{2}$
bands implies a negative value for $f_{\alpha \alpha }-f_{\alpha \alpha '}$.
This result is clearly unphysical.  In addition, $f_{rr}$
from equation 2 for the 2$^{1}$ level, was calculated to be only
0.73 aJ/~\AA$^{2}$.  This value is in sharp disagreement with both the {\it ab
initio} value
and as discussed later, the value obtained by consideration of the observed
zero point energies. We have clearly determined anomalous centrifugal
distortion constants.

Anomalous centrifugal distortion constants often arise when a
vibrational state is perturbed by a nearby level.  In the
present case, no such level appears available.  For the 2$^{1}$ level,
$\tau_{xxxx}$
 appears to be poorly determined.  However, in a fit without distortion
constants,
all the low J transitions have positive residuals while all the high J
transitions have negative residuals.  $D_{J}$ must certainly be negative!
Therefore, the centrifugal distortion constants obtained cannot be used to
derive
physically meaningful harmonic force constants.

     The determination of the harmonic force field had to be
obtained in a less direct manner.  From the observed $\nu_{2}$ vibrational
progressions, we determine the out-of-plane bending force constant, $f_{hh}$.
Since the higher members of the $\nu_{2}$ progression are broad and
erratically spaced\protect\cite{Vaida:1987},  we use only
the data from the 2$^{0}$, 2$^{1}$, and 2$^{2}$ band systems.  From these
levels for each of our isotopomers, $\nu_{2}$ and X$_{22}$ were determined.
These
values are presented in table~\protect\ref{Observed_Freqs}.  Since the X$_{22}$
values are small, the $\nu_{2}$
frequencies are expected to scale as the inverse square of the reduced
mass\protect\cite{Laurie:1958}.
However, as the reduced mass is increased, the effective force constants
systematically
decrease; the force constant obtained from the ND$_{3}$ progression
underestimates the NH$_{3}$ $\nu_{2}$ fundamental frequency by approximately
40~cm$^{-1}$.  We were able to still reasonably estimate $f_{hh}$
at 1.06(4) aJ/~\AA$^{2}$, nearly double the {\it ab initio\/} value of
0.66 aJ/~\AA$^{2}$\protect\cite{Rosmus:1987}.

     Next, the in-plane force constant, $f_{\alpha\alpha }-f_{\alpha \alpha
'}$,
was determined.  We can use the observed inertial defects to estimate this
force
constant.  The inertial defect, $I_{C} - 2I_{B}$, is given for the
2$^{0}$, 2$^{1}$, and 2$^{2}$ vibronic levels for the two
pure isotopomers in table~\protect\ref{Defects}.  We assume that the
centrifugal and
electronic contributions to the inertial defect are small in comparison
to the vibrational contribution and neglect their effect entirely.  Therefore,
the inertial defect was fit to\protect\cite{Oka:1961}:
\begin{equation}
\Delta_{vib}=\sum_{s}\frac{h}{\pi^{2}c}(\nu_{s}+\frac{1}{2})
\sum_{s'}\frac{\omega_{s'}^{2}}{\omega_{s}(\omega_{s}^{2}-\omega_{s'}^{2})}
[(\zeta_{ss'}^{(x)})^{2}+(\zeta_{ss'}^{(z)})^{2}-(\zeta_{ss'}^{(y)})^{2}]
+\sum_{t}\frac{h}{\pi^{2}c}\frac{3}{2\omega_{t}}(\nu_{t}+\frac{1}{2}).
\label{inertial}
\end{equation}
This inertial defect depends principally upon $\nu_{2}$ and
$\nu_{4}$ and only very weakly on the other modes.  Since $\nu_{2}$ has already
been
accurately determined, analysis of the inertial defect should be a sensitive
method to
determine $f_{\alpha\alpha }-f_{\alpha \alpha '}$. Using the {\it ab initio\/}
value for $f_{rr}$ of 5.09 aJ/~\AA$^{2}$\protect\cite{Rosmus:1987}, and our
calculated value for $f_{hh}$,
we fit to the observed inertial defects and determine
$f_{\alpha\alpha }-f_{\alpha \alpha '}$ to be 0.25(2) aJ.
The calculated inertial defects are also included in
table~\protect\ref{Defects}
for comparison.  It should be mentioned that this value for $f_{\alpha \alpha
}-f_{\alpha \alpha '}$
is significantly lower than the ground state value of
0.68 aJ/~\AA$^{2}$\protect\cite{Hargiss:1988}.
  Changing the assumed value of $f_{rr}$ by
{}~20\% changes the calculated inertial defect by $<$0.5\%.

The calculated inertial defects demonstrated systematic
errors from the observed ND$_{3}$ values.  Even changing the $f_{hh}$ force
constant considerably while varying $f_{\alpha\alpha }-f_{\alpha \alpha '}$
could not reproduce the large positive inertial defect we observed for the
ND$_{3}$ 2$^{0}$ level.  In our approach, we have neglected centrifugal
contributions to the inertial defect.  This contribution is given
by\protect\cite{Oka:1961}
$$ \Delta_{cent}=-\hbar^{4}\tau_{zxzx}(\frac{3}{4}\frac{I_{yy}}{C}%
+\frac{I_{xx}}{2B}+\frac{I_{zz}}{2A}).$$
This contribution is predicted to be on the order of $10^{-3}$ and far too
small to explain our large deviation from the observed values.  We suggest that
perturbations in this level are causing $\tau_{zxzx}$ to be anomalously large
such that the centrifugal contributions to the inertial defect are much larger
than predicted.  However, since the observed inertial defects for NH$_{3}$ are
well reporduced as well as the relative spacings for the ND$_{3}$ levels, we
feel that our determined value for $f_{\alpha\alpha }-f_{\alpha \alpha '}$ is
reasonably close to the true value.

Now we turn attention to the perturbation observed in
the NH$_{3}$ 2$^{4}$ band.  It had been suggested that this asymmetry reflected
a
partially resolved structure built off one quantum in the $\nu_{1}$
mode\protect\cite{Vaida:1987};
 thus, the 2$^{4}$ band system
would be overlapped by the 1$^{1}$2$^{1}$ band system.  The 2$^{4}$ and
1$^{1}$2$^{1}$ vibronic levels
have opposite symmetry with respect to reflection in the plane, and thus can
not interact through anharmonic interactions.  Furthermore, the 1$^{1}$2$^{1}$
assignment can be ruled out by the phase of the observed MODR signals.
Transitions
to the 2$^{4}$ level originate from symmetric levels in the
ground electronic state
(leading to an initial emission of the microwave radiation) while transitions
to the 1$^{1}$2$^{1}$ level arise from asymmetric levels (leading to
absorption).  Thus,
the phase of the MODR signal from these two transitions
would have opposite signs.  Since the observed asymmetry in the 2$^{4}$ MODR
spectrum was the same as in absorption, this asymmetry is not the result
of the 1$^{1}$2$^{1}$ band.

The observation of this asymmetry implies a coupling of
the $\nu_{2}$ mode that is greater than its coupling with the dissociative
continuum.  If the mode attributable to this perturbation were either of the
stretching modes, the dissociation rate from the quasibound state would have
been
enhanced greatly; hence, broader lines would have been expected than
what we observed.  The only other mode that could be responsible for
the observed perturbation is due to $\nu_{4}$, the degenerate in-plane bending
mode.  Using the above value for $f_{\alpha\alpha }-f_{\alpha \alpha '}$
 we predict that the harmonic frequency for $\nu_{4}$ is 1114~cm$^{-1}$.
  We assign the observed
perturbation of the 2$^{4}$ level to an anharmonic
mixing with the A$_{1}$ component of the 4$^{3}$ levelnad thus, the two
levels are brought into resonance assuming $\nu_{4}$ = 1114~cm$^{-1}$ and
X$_{44}$ = 25~cm$^{-1}$.  We have neglected
the contribution of g$_{44}$ for lack of data.  The 4$^{3}$ level has
a harmonic amplitude of $\pm$ 43 degrees so we would expect a substantial
positive
anharmonic correction due to the steric repulsion of the hydrogen
atoms.

     We are then left with only one force constant,
$f_{rr}$, to determine.  To estimate this force constant, we use our observed
isotopic
shifts in band origins.  This shift results from changes in vibrational
zero point energy due to isotopic substitution.  We assume the harmonic
approximation and use,
\begin{equation}
T_{o}=T_{e}+E_{o}^{upper}+E_{o}^{lower}
\label{zeropoint}
\end{equation}
where $T_{o}$ is the observed 0-0 band origin for a given
isotopomer, $T_{e}$ is the
energy difference between the two potential surface minima,
and $E_{o}^{upper}$ and $E_{o}^{lower}$ are the
zero point energies for the \~A and \~X states
respectively.  $E_{o}^{lower}$ is taken from the {\it ab initio\/}
work of Hargiss and Ermler\protect\cite{Hargiss:1988}  while $T_{o}$ was known
for each
of our isotopomers.  Constraining $T_{e}$ to be constant and utilizing
our determined values for the other \~A state force constants, $f_{rr}$ was
varied to reproduce the observed band origins for the two pure
isotopomers.  We then used the $T_{o}$ values for the mixed isotopomers as a
consistency check.  Using this procedure, $f_{rr}$ was determined to be 4.2
aJ/~\AA$^{2}$ while $T_{e}$ was found at 48070%
{}~cm$^{-1}$.

The determined value for $f_{rr}$ is approximately 80\%
of the value determined by Rosmus and coworkers\protect\cite{Rosmus:1987}.
We attribute this
discrepancy to a neglect of anharmonicity in the excited state potential.  In
order to estimate these effects, we use the {\it ab initio\/} values for
X$_{NH}$
 = -236~cm$^{-1}$ and X$_{ND}$ = -118~cm$^{-1}$ and
correct for the anharmonicity of the excited state.  After
this correction, $f_{rr}$ was readjusted to 4.9 aJ/~\AA$^{2}$ in better
agreement
with the {it ab initio\/} results\protect\cite{Rosmus:1987}.   Our
estimated values of the force constants, $T_{e}$, and the
vibrational frequencies for all four isotopomers are presented in
table~\protect\ref{Calculated_Freqs}.

To estimate any effect due to the neglect of $f_{\alpha r}-f_{\alpha r'}$
and $f_{rr'}$ on the observed zero point energies, we insert these into the
above
calculations and redetermine our $T_{e}$ and $f_{rr}$ values.  Inclusion of
$f_{rr'}$ at its ground state value of
0.31 aJ/~\AA$^{2}$ changes the calculated $T_{e}$ values by a uniform 14%
{}~cm$^{-1}$.  Likewise, addition of $f_{\alpha r}-f_{\alpha r'}$
at its ground state value of -0.176 aJ increases $T_{e}$ by
a uniform 20~cm$^{-1}$.  Therefore, the neglect of
these two force constants has little effect on our calculated values for
$f_{rr}$
and $T_{e}$.  In addition, increasing $f_{rr}$ by 0.1
aJ/~\AA$^{2}$ changes $T_{e}$(NH$_{3}$) - $T_{e}$(ND$_{3}$) by 14~cm$^{-1}$.
Hence, this method is a sensitive
measure of the $f_{rr}$ force constant.  Given the above
sensitivities to unknown
potential parameters, we suggest that $T_{e}$ = 48070~$\pm$~200~cm$^{-1}$.  The
{\it ab initio\/} predictions for $T_{e}$ were 44260~cm$^{-1}$ (CASSCF) and
45380~cm$^{-1}$ (CEPA) respectively.

The NH$_{3}$ and ND$_{3}$ $\nu_{1}$ frequencies calculated from the derived
harmonic
force field agrees very well with the {\it ab initio\/} determined
values\protect\cite{Rosmus:1987,McCarty:1987}.     These values are about
500~cm$^{-1}$
higher than the values assigned to the fundamentals through dispersed
fluorescence
from the \~C$^{'}$ state\protect\cite{Ashfold:1986a}.  While the anharmonic
corrections to the
fundamental frequency are quite large, the $\nu_{1}$ fundamental is above the
barrier
to dissociation along one N--H bond, and as discussed in the following paper,
we
believe it should be extremely diffuse.  We propose to reassign the broad
features observed by Ashfold {\it et al.}\protect\cite{Ashfold:1986a}  to the
4$^{2}$ band.
On the basis of our above values for $\omega_{4}$
and X$_{44}$, we predict the 4$^{2}$ band should lie near 2425~cm$^{-1}$
compared to 2350~cm$^{-1}$ for the center of the feature previously assigned as
1${^1}$.  The width, and perhaps a downward shift of the 4${^2}$ band, we
believe to
arise from an anharmonic mixing of this level with the continuum created
by the N--H bond dissociation.
{}From our harmonic force field, we
predict that the ND$_{3}$ 4$^{2}$ band should lie at 1730~cm$^{-1}$.
In the ND$_{3}$ dispersed fluorescence spectrum\protect\cite{Ashfold:1986a}, a
broad but highly asymmetric feature was observed at 1790 $\pm$
50~cm$^{-1}$ and
assigned to the 1${^1}$ band very near to our predicted value for
the 4${^2}$ band.

With this newly determined value for $\omega_{4}$, we had hoped that a
progression built from this mode could be directly observed
in the fluorescence excitation spectrum of ND$_{3}$.  While this
level is not Franck-Condon active, the large Coriolis interaction between modes
$\nu_{2}$ and $\nu_{4}$ predicted by the above force field will give intensity
to
the upper J levels of the 4$^{1}$ band.  However, in the expected region, an
overlapping $\nu_{2}$ hot band transition made direct observation of this mode
difficult.
Still an extremely weak and broad transition lying approximately 815~cm$^{-1}$
above the band origin was observed in the wings of the hot band transition that
we
have assigned to the $\nu_{4}$ mode.  It should be noted that the signal to
noise ratio was low and thus the assignment is tentative.

\section{Conclusion}

We have demonstrated the MODR technique used previously to
investigate the vibrational overtones of ammonia can also be
effectively used to investigate the predissociative \~A state electronic
spectrum.  Using this method, we have recorded the electronic spectrum
of NH$_{3}$, NH$_{2}$D, and NHD$_{2}$ with rotational resolution.  Our results
for
NH$_{3}$ represent a significant improvement upon previous experimental
results%
\protect\cite{Douglas:1963,Ashfold:1986a,Xie:1986}  while
the results for the mixed isotopomers represent the first
high resolution studies of these species.  To investigate ND$_{3}$ with
the same resolution as the hydrogen containing species, a strip-line cell was
constructed that allowed the simultaneous passage of radio-frequency and
ultraviolet
radiation.  We have used this cell to record the electronic spectrum of
ND$_{3}$
with improved rotational resolution.  Thus far, these results represent the
most
thorough spectroscopic investigation of the \~A state electronic spectrum.

We have used the rotational constants derived from individual
rovibrational transitions to determine the ammonia \~A state
equilibrium bond length.  We estimate this bond length to be
1.055(8)~\AA.  This
value is significantly different from the previous
experimentally determined
values \protect\cite{Walsh:1961,Douglas:1963} but is quite close to the {\it ab
initio\/}
 value of Rosmus and coworkers\protect\cite{Rosmus:1987}.
In addition, for the first time, the \~A state harmonic
force field has been experimentally estimated.  Where applicable, these
estimates were compared to the {\it ab initio\/} determined
values\protect\cite{Rosmus:1987,McCarty:1987}.
   From the calculated
vibrational term values, we predict $f_{hh}$ is 1.06(4)
aJ/~\AA$^{2}$, nearly double the value determined by Rosmus
{\it et al.}\protect\cite{Rosmus:1987}.
  Using the observed inertial
defects, we predict $f_{\alpha \alpha}-f_{\alpha \alpha '}$ to be 0.25(2) aJ
and
$f_{rr}$ was found to be 4.9 aJ/~\AA$^{2}$ through consideration of the shifts
in band origin upon isotopic substitution.
$f_{rr}$ agrees well with the value determined through the
{\it ab initio\/} work while $f_{\alpha \alpha}-f_{\alpha \alpha '}$ is
approximately 40\% of the
ground state value\protect\cite{Hargiss:1988}.

Strong asymmetry was observed in rovibronic transitions
to the \~A state 2$^{4}$ level.  This asymmetry has been attributed to a
strong anharmonic resonance with the 4$^{3}$ level.  Attempts to
estimate the force constants more directly by analysis of the centrifugal
distortion constants failed.  Anomalous centrifugal distortion constants were
obtained for both the ND$_{3}$ 2$^{0}$ and 2$^{1}$ levels indicating
significant
interaction with a nearby level.

\begin{table}
\begin{center}
\caption{Frequencies of the NH$_{3}$ \~A $\leftarrow$ \~X state transitions
observed by MODR.}
\label{NH3Freqs}
2$^{0}$ band.
\begin{tabular}{ccc}
{\bf Transition}  &  {\bf $\nu_{o}$ (cm$^{-1}$)}  &  {\bf {1$\sigma$}}  \\
\hline
Q$_{2}$(2)  &   46144.370    &  0.037  \\
R$_{2}$(2)  &   46201.910    &  0.115  \\
Q$_{3}$(3)  &   46137.020    &  0.028  \\
R$_{3}$(3)  &   46214.420    &  0.127  \\
Q$_{4}$(4)  &   46127.340    &  0.065  \\
R$_{4}$(4)  &   46225.220    &  0.456  \\
Q$_{4}$(5)  &   46125.220    &  0.147  \\
R$_{4}$(5)  &   46238.290    &  1.782  \\
Q$_{5}$(5)  &   46114.120    &  0.052  \\
R$_{5}$(5)  &   46228.680    &  0.495  \\
Q$_{6}$(6)  &   46098.280    &  0.049  \\
R$_{6}$(6)  &   46230.880    &  0.797  \\
P$_{6}$(7)  &   45958.160    &  0.842  \\
Q$_{6}$(7)  &   46095.960    &  0.078  \\
R$_{6}$(7)  &   46251.720    &  0.590  \\
Q$_{7}$(7)  &   46079.610    &  0.076  \\
R$_{7}$(7)  &   46234.240    &  0.545  \\
Q$_{8}$(8)  &   46058.270    &  0.060  \\
R$_{8}$(8)  &   46233.910    &  1.113  \\
Q$_{9}$(9)  &   46034.510    &  0.224
\end{tabular}
2$^{1}$ band.
\begin{tabular}{ccc}
{\bf Transition}  &  {\bf $\nu_{o}$ (cm$^{-1}$)}  &  {\bf {1$\sigma$}}  \\
\hline
Q$_{1}$(1)    & 47039.410  &  0.044   \\
R$_{1}$(1)    & 47075.670  &  0.093   \\
P$_{2}$(3)    & 46975.220  &  0.095   \\
Q$_{2}$(3)    & 47029.730  &  0.029   \\
R$_{2}$(3)    & 47101.370  &  0.046   \\
Q$_{3}$(3)    & 47028.240  &  0.012   \\
R$_{3}$(3)    & 47099.560  &  0.057   \\
P$_{2}$(5)    & 46921.520  &  0.193   \\
Q$_{2}$(5)    & 47010.530  &  0.274   \\
R$_{2}$(5)    & 47121.470  &  0.158   \\
P$_{4}$(5)    & 46918.020  &  0.091   \\
Q$_{4}$(5)    & 47008.240  &  0.199   \\
R$_{4}$(5)    & 47116.600  &  0.102   \\
Q$_{4}$(5)    & 47007.310  &  0.037   \\
R$_{4}$(5)    & 47113.960  &  0.370   \\
P$_{3}$(6)    & 46890.820  &  0.254   \\
Q$_{3}$(6)    & 46999.900  &  0.189   \\
R$_{3}$(6)    & 47127.040  &  0.206   \\
P$_{5}$(6)    & 46885.750  &  0.483   \\
Q$_{5}$(6)    & 46991.660  &  0.049   \\
R$_{5}$(6)    & 47118.030  &  0.306   \\
R$_{6}$(7)    & 47119.680  &  0.277   \\
Q$_{7}$(8)    & 46954.520  &  0.076   \\
Q$_{8}$(8)    & 46949.130  &  0.018   \\
R$_{8}$(8)    & 47111.080  &  0.409   \\
P$_{8}$(9)    & 46763.110  &  2.098   \\
R$_{8}$(9)    & 47113.430  &  1.158   \\
Q$_{9}$(9)    & 46926.020  &  0.057   \\
R$_{9}$(9)    & 47108.440  &  2.188
\end{tabular}
2$^{2}$ band.
\begin{tabular}{ccc}
{\bf Transition}  &  {\bf $\nu_{o}$ (cm$^{-1}$)}  &  {\bf {1$\sigma$}}  \\
\hline
Q$_{3}$(3) & 47926.520 & 0.140 \\
R$_{3}$(3) & 47991.890 & 0.590 \\
Q$_{6}$(6) & 47887.380 & 0.080 \\
R$_{6}$(6) & 48006.360 & 1.580
\end{tabular}
2$^{3}$ band.
\begin{tabular}{ccc}
{\bf Transition}  &  {\bf $\nu_{o}$ (cm$^{-1}$)}  &  {\bf {1$\sigma$}}  \\
\hline
Q$_{1}$(1) & 48857.500 & 0.260 \\
R$_{1}$(1) & 48890.700 & 0.520 \\
Q$_{2}$(2) & 48848.200 & 0.520 \\
R$_{2}$(2) & 48898.000 & 1.000 \\
Q$_{3}$(3) & 48836.600 & 0.180 \\
R$_{3}$(3) & 48903.000 & 0.360 \\
Q$_{3}$(4) & 48828.500 & 0.680 \\
R$_{3}$(4) & 48911.500 & 1.400 \\
P$_{3}$(4) & 48762.100 & 1.400 \\
Q$_{4}$(4) & 48826.300 & 0.190 \\
R$_{4}$(4) & 48907.600 & 0.380 \\
P$_{3}$(5) & 48721.300 & 0.470 \\
Q$_{3}$(5) & 48805.800 & 0.230 \\
R$_{3}$(5) & 48907.300 & 0.280 \\
Q$_{5}$(5) & 48815.200 & 0.280 \\
R$_{3}$(5) & 48912.800 & 0.560 \\
P$_{3}$(6) & 48689.900 & 1.100 \\
Q$_{3}$(6) & 48788.000 & 0.820 \\
R$_{3}$(6) & 48903.400 & 0.790 \\
Q$_{6}$(6) & 48800.200 & 0.190 \\
R$_{6}$(6) & 48916.400 & 0.380 \\
P$_{6}$(7) & 48671.300 & 0.730 \\
Q$_{6}$(7) & 48775.800 & 0.130 \\
R$_{6}$(7) & 48912.700 & 0.600 \\
Q$_{7}$(7) & 48788.000 & 0.140 \\
R$_{7}$(7) & 48930.200 & 2.400 \\
P$_{6}$(8) & 48616.100 & 0.760 \\
Q$_{6}$(8) & 48748.900 & 0.380 \\
R$_{6}$(8) & 48898.300 & 0.760
\end{tabular}
\end{center}

\end{table}

\begin{table}

\caption{Frequencies of the NH$_{2}$D \~A $\leftarrow$ \~X state transitions
observed by MODR.}
\label{NH2DFreqs}
\begin{center}
2$^{0}$ band.
\begin{tabular}{ccc}
{\bf Transition}  &  {\bf $\nu_{o}$ (cm$^{-1}$)}  &  {\bf {1$\sigma$}}  \\
\hline
3$_{03}$ $\leftarrow$ 3$_{13}$ & 46318.81  & 0.06 \\
4$_{23}$ $\leftarrow$ 3$_{13}$ & 46377.17  & 0.27 \\
4$_{14}$ $\leftarrow$ 4$_{04}$ & 46312.12  & 0.05 \\
5$_{14}$ $\leftarrow$ 4$_{04}$ & 46382.77  & 0.37 \\
5$_{15}$ $\leftarrow$ 5$_{05}$ & 46302.82  & 0.10 \\
6$_{15}$ $\leftarrow$ 5$_{05}$ & 46386.90  & 0.76 \\
6$_{06}$ $\leftarrow$ 7$_{16}$ & 46184.25  & 2.52 \\
7$_{26}$ $\leftarrow$ 7$_{16}$ & 46292.49  & 0.31 \\
7$_{07}$ $\leftarrow$ 8$_{17}$ & 46162.34  & 1.42 \\
8$_{27}$ $\leftarrow$ 8$_{17}$ & 46277.10  & 0.10 \\
9$_{27}$ $\leftarrow$ 8$_{17}$ & 46407.45  & 0.60
\end{tabular}
2$^{1}$ band.
\begin{tabular}{ccc}
{\bf Transition}  &  {\bf $\nu_{o}$ (cm$^{-1}$)}  &  {\bf {1$\sigma$}}  \\
\hline
3$_{13}$ $\leftarrow$ 3$_{03}$ & 47131.70  & 0.02 \\
4$_{13}$ $\leftarrow$ 3$_{03}$ & 47183.77  & 0.13 \\
4$_{04}$ $\leftarrow$ 4$_{14}$ & 47123.14  & 0.10 \\
5$_{24}$ $\leftarrow$ 4$_{14}$ & 47191.81  & 0.97 \\
5$_{05}$ $\leftarrow$ 5$_{15}$ & 47114.68  & 0.06 \\
6$_{25}$ $\leftarrow$ 5$_{15}$ & 47196.05  & 0.97 \\
6$_{16}$ $\leftarrow$ 7$_{26}$ & 46999.00  & 0.60 \\
7$_{16}$ $\leftarrow$ 7$_{26}$ & 47095.97  & 0.08 \\
8$_{36}$ $\leftarrow$ 7$_{26}$ & 47207.14  & 0.38 \\
7$_{17}$ $\leftarrow$ 8$_{27}$ & 46969.76  & 0.44 \\
8$_{17}$ $\leftarrow$ 8$_{27}$ & 47081.86  & 0.05
\end{tabular}
2$^{2}$ band.
\begin{tabular}{ccc}
{\bf Transition}  &  {\bf $\nu_{o}$ (cm$^{-1}$)}  &  {\bf {1$\sigma$}}  \\
\hline
3$_{03}$ $\leftarrow$ 3$_{13}$ & 47958.77  & 0.08 \\
4$_{23}$ $\leftarrow$ 3$_{13}$ & 48016.74  & 0.45 \\
4$_{14}$ $\leftarrow$ 4$_{04}$ & 47952.47  & 0.19 \\
5$_{15}$ $\leftarrow$ 5$_{05}$ & 47942.78  & 0.08 \\
6$_{15}$ $\leftarrow$ 5$_{05}$ & 48025.22  & 1.02 \\
7$_{26}$ $\leftarrow$ 7$_{16}$ & 47918.55  & 0.34 \\
7$_{07}$ $\leftarrow$ 8$_{17}$ & 47806.90  & 1.58 \\
8$_{27}$ $\leftarrow$ 8$_{17}$ & 47904.42  & 0.15 \\
9$_{27}$ $\leftarrow$ 8$_{17}$ & 48033.32  & 1.70
\end{tabular}
\end{center}

\end{table}

\begin{table}
\begin{center}

\caption{Frequencies of the NHD$_{2}$ \~A $\leftarrow$ \~X state transitions
observed by MODR.}
\label{NHD2Freqs}
2$^{0}$ band.
\begin{tabular}{ccc}
{\bf Transition}  &  {\bf $\nu_{o}$ (cm$^{-1}$)}  &  {\bf {1$\sigma$}}  \\
\hline
2$_{02}$ $\leftarrow$ 2$_{12}$ & 46508.41  & 0.07 \\
3$_{22}$ $\leftarrow$ 2$_{12}$ & 46546.54  & 0.21 \\
3$_{13}$ $\leftarrow$ 3$_{03}$ & 46505.93  & 0.10 \\
4$_{13}$ $\leftarrow$ 3$_{03}$ & 46550.36  & 0.50 \\
4$_{14}$ $\leftarrow$ 5$_{24}$ & 46438.30  & 0.48 \\
5$_{14}$ $\leftarrow$ 5$_{24}$ & 46496.92  & 0.07 \\
6$_{34}$ $\leftarrow$ 5$_{24}$ & 46569.67  & 0.23
\end{tabular}
2$^{1}$ band.
\begin{tabular}{ccc}
{\bf Transition}  &  {\bf $\nu_{o}$ (cm$^{-1}$)}  &  {\bf {1$\sigma$}}  \\
\hline
2$_{12}$ $\leftarrow$ 2$_{02}$ & 47248.53  & 0.02 \\
3$_{12}$ $\leftarrow$ 2$_{02}$ & 47279.38  & 0.07 \\
3$_{03}$ $\leftarrow$ 3$_{13}$ & 47242.99  & 0.04 \\
4$_{23}$ $\leftarrow$ 3$_{13}$ & 47288.89  & 0.19 \\
4$_{04}$ $\leftarrow$ 5$_{14}$ & 47176.79  & 0.11 \\
5$_{24}$ $\leftarrow$ 5$_{14}$ & 47233.86  & 0.02 \\
6$_{24}$ $\leftarrow$ 5$_{14}$ & 47297.90  & 0.05
\end{tabular}
2$^{2}$ band.
\begin{tabular}{ccc}
{\bf Transition}  &  {\bf $\nu_{o}$ (cm$^{-1}$)}  &  {\bf {1$\sigma$}}  \\
\hline
2$_{02}$ $\leftarrow$ 2$_{12}$ & 47994.52  & 0.13 \\
3$_{22}$ $\leftarrow$ 2$_{12}$ & 48028.42  & 0.40 \\
3$_{13}$ $\leftarrow$ 3$_{03}$ & 47991.77  & 0.19 \\
4$_{13}$ $\leftarrow$ 3$_{03}$ & 48035.52  & 0.92 \\
4$_{14}$ $\leftarrow$ 5$_{24}$ & 47923.51  & 0.45 \\
5$_{14}$ $\leftarrow$ 5$_{24}$ & 47977.84  & 0.06 \\
6$_{34}$ $\leftarrow$ 5$_{24}$ & 48042.56  & 0.20
\end{tabular}
\end{center}
\end{table}

\begin{table}
\begin{center}
\caption{Frequencies of the ND$_{3}$ \~A $\leftarrow$ \~X state
transitions observed by MODR.}
\label{ND3Freqs}
2$^{0}$ band.
\begin{tabular}{ccc}
{\bf Transition}  &  {\bf $\nu_{o}$ (cm$^{-1}$)}  &  {\bf {1$\sigma$}}  \\
\hline
Q$_{3}$(3) & 46706.933 &  0.0442 \\
Q$_{4}$(4) & 46702.008 &  0.0430 \\
Q$_{3}$(5) & 46703.979 &  0.0558 \\
Q$_{5}$(5) & 46695.713 &  0.0190 \\
Q$_{6}$(6) & 46687.893 &  0.0229 \\
Q$_{6}$(7) & 46685.503 &  0.0444 \\
Q$_{7}$(7) & 46679.039 &  0.0172 \\
Q$_{6}$(8) & 46682.565 &  0.0248 \\
Q$_{7}$(8) & 46675.908 &  0.0236 \\
Q$_{8}$(8) & 46668.619 &  0.0117 \\
Q$_{6}$(9) & 46675.608 &  0.1147 \\
Q$_{9}$(9) & 46656.743 &  0.0312 \\
Q$_{9}$(10)& 46653.319 &  0.0293
\end{tabular}
2$^{1}$ band.
\begin{tabular}{ccc}
{\bf Transition}  &  {\bf $\nu_{o}$ (cm$^{-1}$)}  &  {\bf {1$\sigma$}}  \\
\hline
Q$_{1}$(1)  &    47366.436  & 0.0204 \\
Q$_{2}$(2)  &    47364.104  & 0.0267 \\
Q$_{3}$(3)  &    47360.481  & 0.0095 \\
R$_{3}$(3)  &    47360.481  & 0.0135 \\
Q$_{3}$(4)  &    47357.568  & 0.0142 \\
Q$_{4}$(4)  &    47355.529  & 0.0042 \\
R$_{4}$(4)  &    47403.190  & 0.0319 \\
Q$_{3}$(5)  &    47353.913  & 0.0264 \\
Q$_{5}$(5)  &    47349.359  & 0.0164 \\
Q$_{3}$(6)  &    47349.620  & 0.0253 \\
Q$_{5}$(6)  &    47344.960  & 0.0165 \\
Q$_{6}$(6)  &    47341.996  & 0.0086 \\
Q$_{6}$(7)  &    47336.624  & 0.0091 \\
Q$_{7}$(7)  &    47332.976  & 0.0066 \\
Q$_{6}$(8)  &    47331.091  & 0.0103 \\
Q$_{7}$(8)  &    47327.133  & 0.0256 \\
Q$_{8}$(8)  &    47322.896  & 0.0057 \\
Q$_{3}$(9)  &    47333.513  & 0.0657 \\
Q$_{6}$(9)  &    47324.945  & 0.0468 \\
Q$_{9}$(9)  &    47311.590  & 0.0155 \\
Q$_{9}$(10) &    47304.254  & 0.0129 \\
Q$_{9}$(11) &    47295.952  & 0.0235 \\
Q$_{10}$(11)&    47290.568  & 0.0273 \\
Q$_{11}$(12)&    47275.774  & 0.0225 \\
Q$_{12}$(14)&    47249.702  & 0.0382
\end{tabular}
2$^{2}$ band.
\begin{tabular}{ccc}
{\bf Transition}  &  {\bf $\nu_{o}$ (cm$^{-1}$)}  &  {\bf {1$\sigma$}}  \\
\hline
Q$_{3}$(3) &       48024.337 &   0.0528 \\
Q$_{3}$(4) &       48019.452 &   0.1298 \\
Q$_{4}$(4) &       48019.761 &   0.0940 \\
Q$_{3}$(5) &       48014.317 &   0.1770 \\
Q$_{5}$(5) &       48013.202 &   0.0768 \\
Q$_{6}$(6) &       48005.929 &   0.0704 \\
Q$_{6}$(7) &       47997.295 &   0.2152 \\
Q$_{7}$(8) &       47988.188 &   0.1127 \\
Q$_{9}$(9) &       47975.291 &   0.2514
\end{tabular}
\end{center}
\end{table}

\begin{table}
\begin{center}
\caption{Constants derived from rotational band fits of the
\~A $\leftarrow$ \~X transitions in NH$_{3}$, NH$_{2}$D, and NHD$_{2}$.}
\label{HFits}
\begin{tabular}{cddddd}
& {\bf $\nu_{2}$} & {\bf $\nu_{o}$(cm$^{-1}$)} & {\bf A(cm$^{-1}$)} &
{\bf B(cm$^{-1}$)} & {\bf C(cm$^{-1}$)}\\
{\bf For NH$_{3}$.} & 0 & 46150(1) & & 9.71(6) & 4.80(2) \\
 & & & & 9.72$^{\dag}$ & 4.79$^{\dag}$ \\
 & & & & 9.6$^{\ddag}$  & 4.8$^{\ddag}$  \\
 & 1 & 47042.2(8) & & 9.01(3) & 4.87(2) \\
 & & & & 9.05$^{\dag}$ & 4.85$^{\dag}$ \\
 & & & & 9.05$^{\ddag}$ & 5.2$^{\ddag}$  \\
 & 2 & 47942(3) & & 8.4(3) & 5.0(1) \\
 & & & & 8.66$^{\dag}$ & 4.87$^{\dag}$ \\
  & & & & 8.5$^{\ddag}$  & 5.4$^{\ddag}$ \\
 & 3 & 48856(2) & & 8.21(8) & 4.98(8) \\
 & & & & 8.4$^{b}$ & 5.8$^{\dag}$ \\
 & 4 & 49978(5) & & & \\
 & & & & & \\ \hline
{\bf For NH$_{2}$D.}  & 0 & 46328.9(3) & 8.9 & 6.37(30) & 3.649(9) \\
 & 1 & 47141.38(13) & 9.1 & 5.78(6) & 3.737(5) \\
 & 2 & 47970.4(5)  & 7.8   & 5.9(6) & 3.782(20) \\
 & & & & & \\ \hline
{\bf For NHD$_{2}$.} & 0 & 46513.14(14) & 7.3 & 5.06(4) & 2.936(5) \\
 & 1 & 47251.58(9) & 7.0& 4.809(13) & 2.987(3) \\
 & 2 & 47999.80(20) & 6.2 & 4.88(10)  & 3.028(8)
\end{tabular}
\end{center}
$^{\dag}$Reference\ \protect\cite{Xie:1986}\\
$^{\ddag}$Reference\ \protect\cite{Ziegler:1985}
\end{table}

\begin{table}
\begin{center}
\caption{Constants derived from rotational band fits of
the ND$_{3}$ \~A $\leftarrow$ \~X transitions.}
\label{DFits}
\begin{tabular}{clddd}
 & {\bf Constant}  & {\bf Fit I}     & {\bf Fit II}   &  {\bf Fit III} \\
{\bf 2$^{0}$ band.} &  {\bf $\nu_{o}$} & 46713.6(2) & 46713.3(1) & \\
 & {\bf B} & 5.02(2) & 5.04(1) & \\
 & {\bf C} & 2.431(9) & 2.438(2) & \\
 &  {\bf $D_{J}$ (x 10$^{4}$)} & 26.(3) & & \\
 &  {\bf $D_{JK}$ (x 10$^{4}$)} & -57.(5) & & \\
 &  {\bf $D_{K}$ (x 10$^{4}$)} & 30.(3) & & \\
 & {\bf $\tau_{xxxx}$ (x 10$^{4}$)} & & -109.(10) & \\
 & {\bf $\tau_{zzzz}$ (x 10$^{4}$)} & & 0$^{a}$ & \\
 & {\bf $\sigma_{err}$} & 0.18 & 0.17 & \\
 & & & & \\ \hline
 {\bf 2$^{1}$ band.} &  {\bf $\nu_{o}$} & 47367.51(5) & 47357.50(5) &
47367.60(9) \\
 & {\bf B} & 4.772(4) & 4.768(4)  & 4.785(8) \\
 & {\bf C} & 2.4745(19) & 2.4779(17)  & 2.465(1) \\
 &  {\bf $D_{J}$ (x 10$^{4}$)} &-2.0(7) & & \\
 &  {\bf $D_{JK}$ (x 10$^{4}$)} & 7.0(15) & & \\
 &  {\bf $D_{K}$ (x 10$^{4}$)} &-4.9(9) & & \\
 & {\bf $\tau_{xxxx}$ (x 10$^{4}$)} & & 8.0(29) & -12.(5)\\
 & {\bf $\tau_{zzzz}$ (x 10$^{4}$)} & & -2.09(24) & 0.$^{a}$ \\
 & {\bf $\sigma_{err}$} & 0.1 & 0.12 & 0.21 \\
 & & & & \\ \hline
 {\bf 2$^{2}$ band.} &  {\bf $\nu_{o}$} & 48031.9(2) & 48032.2(4) &  \\
 & {\bf B} & 4.57(2) &  4.40(8) & \\
 & {\bf C} & 2.489(6) & 2.52(2) & \\
 &  {\bf $D_{J}$ (x 10$^{4}$)} & 0.$^{a}$ & & \\
 &  {\bf $D_{JK}$ (x 10$^{4}$)} & 0.$^{a}$ & & \\
 &  {\bf $D_{K}$ (x 10$^{4}$)} & 0.$^{a}$ & & \\
 & {\bf $\tau_{xxxx}$ (x 10$^{4}$)} & & 295.(120) & \\
 & {\bf $\tau_{zzzz}$ (x 10$^{4}$)} & & -10.(7) & \\
 & {\bf $\sigma_{err}$} & 0.33 & 0.23 &
\end{tabular}
\end{center}

Numbers in parentheses are $\pm 1\sigma$.\\
All values given are in wavenumbers.\\$^{a}$Constrained value.\\
$\sigma_{err}$ represents the systematic error.  It is added in
quadrature to the statistical error of each fitted line to make
$\chi^{2}$ for the fit equal to the number of degrees of freedom.

\end{table}

\begin{table}
\caption{Observed and calculated inertial defects for the two `pure'
isotopomers, NH$_{3}$ and ND$_{3}$.  Numbers in parantheses are $\pm 2\sigma$.}
\label{Defects}
\begin{center}
\begin{tabular}{rdd}
\multicolumn{1}{l}{{\bf For ND$_{3}$.}} & {\bf Observed (amu-\AA$^{2}$)} &
  {\bf Calculated (amu-\AA$^{2}$)} \\ \hline \hline
2$^{0}$ band &  +0.218(25) &   +0.083  \\
2$^{1}$ band & -0.264(9) &    -0.326   \\
2$^{2}$ band & -0.61(3) & -0.736  \\
\end{tabular}
\vskip0.2in
$\Delta_{obs} = 0.39(3) - 0.42(2) (\nu_{2} + \frac{1}{2})$
\vskip0.2in
\begin{tabular}{rdd}
\multicolumn{1}{l}{{\bf For NH$_{3}$.}} & {\bf Observed (amu-\AA$^{2}$)} &
  {\bf Calculated (amu-\AA$^{2}$)} \\ \hline \hline
2$^{0}$ band &  +0.04(2) &   +0.064  \\
2$^{1}$ band & -0.28(2) &    -0.232  \\
2$^{2}$ band & -0.6(2) & -0.511 \\
\end{tabular}
\vskip0.2in
     $\Delta_{obs} = 0.19(3) - 0.30(3) (\nu_{2} + \frac{1}{2})$
\vskip0.2in
\end{center}

\end{table}

\begin{table}

\begin{center}

\caption{Out-of-plane vibrational frequencies and anharmonic
constants.}
\label{Observed_Freqs}

\begin{tabular}{ccdd}
& Transition & $\nu_{2}$(cm$^{-1}$) & 2$\chi_{22}$(cm$^{-1}$)\\ \hline
{\bf For NH$_{3}$.} & & & \\
 & 2$^{1}$ $\leftarrow$ 2$^{0}$ & 892. & \\
 & 2$^{2}$ $\leftarrow$ 2$^{1}$ & 900. & 8. \\
 & 2$^{3}$ $\leftarrow$ 2$^{2}$ & 914. & 14. \\
 & 2$^{4}$ $\leftarrow$ 2$^{3}$ & 942. & 28. \\ \hline
 & & & \\
{\bf For NH$_{2}$D.} & & & \\
 & 2$^{1}$ $\leftarrow$ 2$^{0}$ & 812.9 & \\
 & 2$^{2}$ $\leftarrow$ 2$^{1}$ & 830.1 & 17.2 \\
 & 2$^{3}$ $\leftarrow$ 2$^{2}$ & 855.  & 24.9 \\ \hline
& & & \\
{\bf For NHD$_{2}$.} & & & \\
 & 2$^{1}$ $\leftarrow$ 2$^{0}$ & 738.3 & \\
 & 2$^{2}$ $\leftarrow$ 2$^{1}$ & 748.4 & 10.1 \\
 & 2$^{3}$ $\leftarrow$ 2$^{2}$ & 763.5 & 15.1 \\ \hline
& & & \\
{\bf For ND$_{3}$.} & & & \\
 & 2$^{1}$ $\leftarrow$ 2$^{0}$ & & \\
 & 2$^{1}$ $\leftarrow$ 2$^{0}$ & & 10.5
\end{tabular}
\end{center}
\end{table}

\begin{table}

\caption{Calculated \~A state normal mode frequencies derived
from the experimentally determined force field.  All values are given in
cm$^{-1}$.
Values given in parentheses are worst case estimates.}

\begin{center}

\begin{tabular}{cdddd}
{\bf Mode} & {\bf NH$_{3}$} & {\bf NH$_{2}$D} & {\bf NHD$_{2}$} & {\bf
ND$_{3}$} \\
$\nu_{1}$ & 2870.(30) & 2930.(30) & 2980.(30) & 2030.(20) \\
$\nu_{2}$ &  892.(4) &  813.(50)   &  738.(20)  &  653.(10)  \\
$\nu_{3}$ & 3020(30)& 3020.(30) & 2094.(21) & 2244.(23) \\
          &        & 2165.(25) & 2245.(24) &         \\
$\nu_{4}$ & 1110.(50)& 1110.(50) & 1020.(50) &  820.(40) \\
          &        &  930.(40) &  820.(40) &\\
& & & &\\
& \multicolumn{4}{c}{{\bf Zero Point Energy}}  \\
 & 5840.(200)   & 5340.(180)   & 4830.(180)   & 4310.(170) \\
& & & &\\
& \multicolumn{4}{c}{{\bf T$_{e}$}} \\
 & 48071.30  & 48070.63  & 48064.96  & 48070.53 \\
\end{tabular}

\end{center}

\label{Calculated_Freqs}

\end{table}

\begin{figure}
\caption{
Schematic diagram of the microwave detected, microwave-
optical double resonance spectrometer.}
\label{Schematic}
\end{figure}

\begin{figure}
\caption{
Microwave absorption observed from the NH$_{2}$D 3$_{13}$, 0
$\leftarrow$ 3$_{03}$, 1 inversion level following pumping of the lower level
by a
resonant ultraviolet pulse.}
\label{Nutation}
\end{figure}

\begin{figure}

\caption{Cross section of the RFODR cell.  The strip-line is held at
a fixed distance from the ground plane by specially machined KEL-F
blocks.  The laser is passed between the strip-line and the
ground plane.}
\label{Strip_Line}

\end{figure}

\end{document}